\documentclass[8pt]{article}
\usepackage{amsmath,amssymb,amsfonts,euscript}
\usepackage{algorithmic,algorithm}

\usepackage{graphicx}
\usepackage[colorlinks]{hyperref}

\newtheorem{prethm}{{\bf Theorem}}

\newenvironment{thm}{\begin{prethm}{\hspace{-0.5
               em}{\bf}}}{\end{prethm}}

\newtheorem{prepro}{{\bf Theorem}}

\newenvironment{pro}{\begin{prepro}{\hspace{-0.5
               em}{\bf}}}{\end{prepro}}

\newtheorem{preprop}{{\bf Proposition}}

\newtheorem{precor}{{\bf Corollary}}

\newenvironment{cor}{\begin{precor}{\hspace{-0.5
               em}{\bf}}}{\end{precor}}

\newtheorem{preconj}{{\bf Conjecture}}

\newenvironment{conj}{\begin{preconj}{\hspace{-0.5
               em}{\bf}}}{\end{preconj}}

\newtheorem{preconja}{{\bf Conjecture}}

\newtheorem{predefi}{{\bf Definition}}

\newtheorem{preprob}{{\bf Problem}}

\newenvironment{prob}{\begin{preprob}{\hspace{-0.5
               em}{\bf.}}}{\end{preprob}}

\newtheorem{preremark}{{\bf Remark}}

\newenvironment{remark}{\begin{preremark}\rm{\hspace{-0.5
               em}{\bf}}}{\end{preremark}}

\newtheorem{preexample}{{\bf Example}}

\newtheorem{prelem}{{\bf Lemma}}

\newenvironment{lem}{\begin{prelem}{\hspace{-0.5
               em}{\bf}}}{\end{prelem}}

\newtheorem{prelam}{{\bf Problem}}

\newtheorem{preproof}{{\bf Proof}}

\newenvironment{proof}[1]{\begin{preproof}{\rm
               #1}\hfill{$\Box$}}{\end{preproof}}

\newtheorem{preali}{{\bf Proof of Theorem 2.}}

\title{Not-All-Equal and 1-in-Degree Decompositions: Algorithmic Complexity and Applications}

\author{{\normalsize
{  Ali Dehghan${}^{\mathsf{a,b}}$},\,
{  Mohammad-Reza Sadeghi${}^{\mathsf{b}}$},\,
 {  Arash Ahadi${}^{\mathsf{c}}$}\,
}\vspace{3mm}
\\{\footnotesize{${}^{\mathsf{a}}$\it Systems and Computer Engineering Department, Carleton University, Ottawa,   Canada}}
\\{\footnotesize{${}^{\mathsf{b}}$\it Department of Mathematics and Computer Science,
Amirkabir University of Technology, Tehran,
Iran}}  {\footnotesize{}}\\{\footnotesize{${}^{\mathsf{c}}$\it
Department of
Mathematical Sciences, Sharif University of Technology, Tehran, Iran}}
\thanks{{\it E-mail addresses}:  $\mathsf{alidehghan@sce.carleton.ca}$, $\mathsf{msadeghi@aut.ac.ir}$, $\mathsf{arash\_ahadi@mehr.sharif.edu}$. } }

\date{
}

\begin{document}
\maketitle

\begin{abstract}
{\small \noindent
A Not-All-Equal (NAE) decomposition of a graph $G$ is a decomposition  of the vertices of $G$  into
two parts such that each vertex in $G$
has at least one
neighbor in each part. Also, a 1-in-Degree decomposition of a graph $G$ is a decomposition of the vertices of $G$
 into two parts $A$ and $B$
such that each vertex in the graph $G$
has exactly one
neighbor in part $A$.
Among our results, we show that for a given graph $G$, if   $G$
does not have any cycle of length congruent to 2 mod 4, then there is a polynomial time algorithm to decide whether
$G$ has a 1-in-Degree decomposition. In sharp contrast, we prove that for every $r$, $r\geq 3$, for a given $r$-regular bipartite
 graph $G$
determining whether $G$ has a 1-in-Degree decomposition is  $ \mathbf{NP} $-complete.
These complexity results have been especially useful in proving $ \mathbf{NP} $-completeness
of various graph related problems for restricted classes of graphs.
In consequence of these results we show that for a given bipartite 3-regular graph $G$ determining
whether there is a vector in the null-space of the 0,1-adjacency matrix of $G$ such that its
entries belong to $\{ \pm 1,\pm 2\}$ is $\mathbf{NP} $-complete.
Among other results, we introduce a new version of { Planar 1-in-3 SAT} and we
prove that this version is also $ \mathbf{NP} $-complete. In consequence of this result,
we show that for a given planar $(3,4)$-semiregular graph $G$ determining whether there
is a vector in the null-space of the 0,1-incidence matrix of $G$ such that its entries
belong to $\{ \pm 1,\pm 2\}$ is $\mathbf{NP} $-complete.
}

\begin{flushleft}

\end{flushleft}

\end{abstract}


\section{Introduction}
\label{Section1}

Studying  the structure of the null-space of the incidence matrix of any undirected graph is an active
area in linear algebra and computer science. For instance,  Villarreal
\cite{null2} proved that the null-space of
the incidence matrix of every graph has a basis whose elements have entries in $\{-2,-1,0,1,2\}$
(for more information see \cite{hazama2002kernels, sander2005simply,   sander2009tree, sander2009sudoku}).
A {\it zero-sum $k$-flow} for a graph $G$ is a vector in the null-space of the 0,1-incidence matrix
of $G$ such that its entries belong to $\{\pm 1,\ldots,\pm(k-1)\}$.
Recently, zero-sum flows have been studied extensively by several authors, for
example see \cite{gcom1, gcom2,  dehghan2014complexity, sarkis2015zero, NEW5, NEW4, wang2014zero, wang2013zero}.
Zero-sum flows are interesting to study, because of their
connections to Bouchet's 6-flow conjecture (every bidirected graph that has a nowhere-zero bidirected
flow admits a nowhere-zero bidirected 6-flow \cite{AA8}) and Tutte's 5-flow conjecture (every bridgeless graph has a nowhere-zero 5-flow \cite{AA5}). For more information about these two conjectures, see \cite{ mavcajova2014nowhere, thomassen2014group, wei2014nowhere, BB3, BB4, BB1}.
Recently, it was shown that it is $ \mathbf{NP} $-complete to determine whether a given
graph $G$ has a zero-sum 3-flow \cite{dehghan2014complexity}. In this work, we improve the previous hardness result and show that  it is $ \mathbf{NP} $-complete to determine whether a given planar
graph $G$ has a zero-sum 3-flow.

On the other hand, eigenspaces of the adjacency matrix of graphs have been studied
for many years \cite{ MR3523349, sander2010eigenvalues, sander2009tree}.
This is especially the case for the null-space of the adjacency matrix of graphs, which has been studied for a
number of graph classes.
A
simply structured basis is
a basis vector
that contains only entries from the set $\{-1, 0, 1\}$. Simply structured bases have been shown to exist for a number of graph
classes. Mainly, attention is restricted to the graph kernel. The
existing literature features results on trees \cite{sander2005simply}, line graphs of trees \cite{marino2006more, sciriha1999two}, unicyclic
graphs \cite{nath2007null, sander2007simply}, bipartite graphs \cite{cvetkovic1972algebraic}, and cographs \cite{sander2}.
A {\it zero-sum  vertex $k$-flow} is a vector in the null-space of the 0,1-adjacency matrix
of $G$ such that its entries belong to $\{\pm 1,\ldots,\pm (k-1)\}$.
In this paper, we show that for a given 3-regular bipartite graph $G$ determining whether $G$ has
a zero-sum vertex $3$-flow is $ \mathbf{NP} $-complete.

In this work, in order to study zero-sum vertex  flows, we introduce the  concept of Not-All-Equal (NAE) decomposition
and the  concept of 1-in-Degree decomposition
of graphs.
A graph $G$ has a NAE decomposition  if the vertices of the graph
$G$ can be partitioned into two total
dominating sets. In other words, a NAE decomposition of a graph $G$ is a
decomposition of the vertices of $G$ into two parts such that each vertex in the graph $G$
has at least one
neighbor in each part.
Also,  a graph $G$ has a 1-in-Degree decomposition  if the graph $G$ has a total perfect dominating set.
 In other words, a 1-in-Degree decomposition of a graph $G$ is a decomposition of the vertices of $G$ into two sets $A$ and $B$
such that each vertex in the graph $G$
has exactly one
neighbor in part $A$.
The computational complexity of determining  whether a given graph $G$ has a total perfect dominating set
has been studied  by several authors, for instance see
 \cite{brandstadt2015polynomial, brandstadt2012efficient, cattaneo2014parameterized, golovach2012parameterized, HT, HT2}.
 It  was shown that the problem of deciding whether a planar bipartite graph of
maximum degree three has any total perfect dominating set
is $ \mathbf{NP} $-complete \cite{HT2}.
On the other hand, about the partitioning of a graph into two total dominating sets, Zelinka   \cite{ID4} showed
that the large minimum degree is not sufficient to
guarantee the existence of a partition of a graph into two total dominating sets.
Moreover, Calkin and Dankelmann  \cite{HT3} and
Feige {\it et al.}   \cite{HT4} showed that if the maximum degree is not too large relative to the
 minimum degree, then sufficiently large minimum degree does suffice.

\subsection{Our Results}

{\it Planar 1-in-3 SAT}\, is a well-known $ \mathbf{NP} $-complete problem in computational complexity.
\\ \\
 {\it  Planar 1-in-3 SAT.}\\
\textsc{Instance}: A 3SAT formula $\Psi=(X,C)$
such that  the
bipartite graph obtained by linking a variable and a clause if and only
 if the
 variable appears in the clause, is planar.\\
\textsc{Question}: Is there a truth assignment for $X$ such that
 each clause in $C$ has exactly
one { true} literal?

Also, it is well-known that {\it Planar 1-in-3 SAT} is $ \mathbf{NP} $-complete, even
if we add the {\it backbone} between the variable vertices \cite{mulzer2008minimum}. In this work,  we introduce
a new practical version of {\it Planar 1-in-3 SAT} which has a ``tree"  between the clause vertices and we prove that
this version is also $ \mathbf{NP} $-complete. We call this problem  {\it   Monotone Planar Tree-like 1-in-3 SAT}.
\\ \\
 {\it   Monotone Planar Tree-like 1-in-3 SAT.}\\
\textsc{Instance}: A 3SAT formula $\Psi=(X,C)$
such that \\
(i) every variable
appears in exactly three clauses, \\
(ii)  there
is no negation in the formula, and\\
(iii)  the following graph  obtained from $\Psi$ is planar.\\
 The graph has one vertex for each variable,
 one vertex for each clause,
 each clause vertex is
connected by an edge to the variable
vertices corresponding to the literals present in the clause, and
 some clause vertices are connected  to each other such that
the induced subgraph on the set of clause vertices  forms a tree.\\
\textsc{Question}: Is there a truth assignment for $X$ such that
 each clause in $C$ has exactly
one { true} literal?

In order to show that the above problem is $ \mathbf{NP} $-complete, we reduce the following problem to our problem.
Moore and Robson  \cite{MR1863810} proved
that the following version of the 1-in-3 SAT problem is $ \mathbf{NP}$-complete.
\\ \\
 {\it Cubic Planar 1-in-3 SAT.}\\
\textsc{Instance}: A 3SAT formula $\Psi=(X,C)$
such that each variable
appears in exactly three clauses, there
is no negation in the formula, and the
bipartite graph obtained by linking a variable and a clause if and only
 if the
 variable appears in the clause, is planar.\\
\textsc{Question}: Is there a truth assignment for $X$ such that
 each clause in $C$ has exactly
one { true} literal?

Next, we focus on an application of this hardness result.
A {\it zero-sum $k$-flow} for a graph $G$ is a vector in the null-space of the 0,1-incidence matrix
of $G$ such that its entries belong to $\{\pm 1,\ldots,±(k-1)\}$. In other words,
a  zero-sum $k$-flow for a graph $G$ is a labeling of its edges
from the set $\{\pm 1,\ldots,\pm(k-1)\}$
such that the sum of the labels of all edges incident with
each vertex is zero. In 2009, Akbari {\it et al.}  posed the following interesting conjecture about the zero-sum flows \cite{AA1}.

\begin{conj}\noindent {\bf [Zero-Sum Conjecture (ZSC) \rm \cite{AA1}\bf]}
If $G$ is a graph with a zero-sum flow, then the graph $G$ admits
a zero-sum 6-flow.
\end{conj}

In 2010, Akbari {\it et al.} proved that Bouchet's Conjecture (every bidirected graph that has a nowhere-zero bidirected
flow admits a nowhere-zero bidirected 6-flow \cite{AA8}) and Zero-Sum Conjecture are equivalent \cite{AA2}.
Regarding the computational complexity of zero-sum flows, it was shown that it is $ \mathbf{NP} $-complete
 to determine whether a given
$(3,4)$-semiregular graph has a zero-sum 3-flow \cite{dehghan2014complexity}. Here, we improve the previous
complexity result for the class of planar graphs.
We show that for a
given planar $(3,4)$-semiregular graph $G$ determining whether there is a
vector in the null-space of the 0,1-incidence matrix of $G$ such that its
entries belong to $\{ \pm 1,\pm 2\}$ is $\mathbf{NP} $-complete.

$\begin{aligned}
 \text{\it   1-in-3 SAT}               & \leq^p_m   \text{\it   Cubic Planar  1-in-3 SAT} \\
                                       & \leq^p_m    \text{\it   Monotone Planar Tree-like 1-in-3 SAT} \\
                                       &\leq^p_m    \text{\it Finding a Zero-Sum 3-Flow}.
\end{aligned}$

A graph $G$ has a NAE decomposition  if the vertices of the graph
$G$ can be partitioned into two total
dominating sets. In other words, a NAE decomposition of a graph $G$ is a
decomposition of the vertices into two parts such that each vertex in the graph $G$
has at least one
neighbor in each part.
Also,  a graph $G$ has a 1-in-Degree decomposition  if the graph $G$ has a total perfect dominating set.
 In other words, a 1-in-Degree decomposition of a graph $G$ is a decomposition of the vertices into two sets $A$ and $B$
such that each vertex in the graph $G$
has exactly one
neighbor in part $A$.
Since many problems in graph theory and computer science  have the symmetric
structures, NAE and 1-in-Degree decompositions have many applications to prove the $ \mathbf{NP} $-completeness
of the problems.

Next, we consider the computational complexity of the determining whether a graph has a
1-in-Degree decomposition. We prove that
for every $r\geq 3$, for a given $r$-regular bipartite graph $G$
which  $G$ has a   Not-All-Equal decomposition determining
whether $G$ has a 1-in-Degree decomposition is  $ \mathbf{NP} $-complete.
Moreover, we show that  if $G$ is a bipartite graph   and does not have any  cycle of length
congruent to 2 mod 4, then there is a polynomial time algorithm to decide
whether the graph $G$ has a 1-in-Degree decomposition.

On the other hand, regarding the NAE decomposition, we show that if
$G $ is an $r$-regular bipartite graph and $r\geq 4$, then  the graph $G$ always has a NAE decomposition.
A summary of results on the NAE and 1-in-Degree decompositions is shown in Table 1.

\begin{table}[ht]
\begin{center}
\caption{Summary of results on the existence of NAE and 1-in-Degree decompositions}
  \begin{tabular}{ | l | c | c | }
    \hline
    $r$-regular  & The existence of    & The existence of   \\
    bipartite graph &  1-in-Degree decomposition     &       NAE   decomposition  \\   \hline \hline
    $r=3$ & $ \mathbf{NP} $-complete & Open \\ \hline
    $r\geq 4$ & $ \mathbf{NP} $-complete & $ \mathbf{P} $ \\
    \hline
  \end{tabular}
\end{center}
\end{table}

As we can see in Table 1, for the case $r=3$ the complexity of the existence of NAE   decomposition is unknown.
Thus, the computational complexities of the following two interesting problems remain unsolved.
\\ \\
 {\it Cubic Bipartite  NAE Decomposition Problem.}\\
\textsc{Instance}: A 3-regular bipartite graph $G$.\\
\textsc{Question}: Is there a  NAE decomposition for $G$?
\\
\\
 {\it Cubic Monotone  NAE 3SAT.}\\
\textsc{Instance}: Set $X$ of variables, collection $C$ of clauses over $X$ such that for each
clause $c \in C$ we have $\mid c  \mid =3$, every variable appears in
exactly three clauses and there is no negation in the formula.\\
\textsc{Question}: Is there a truth assignment for $X$ such that each clause in $C$ has at
least one {  true} literal and at least one {  false} literal?\\

Note that there is a simple polynomial time reduction from {\it Cubic Bipartite  NAE Decomposition Problem}
 to {\it Cubic Monotone  NAE 3SAT}.
Thus if {\it Cubic Bipartite  NAE Decomposition Problem } is
$ \mathbf{NP} $-complete, then,  {\it Cubic Monotone  NAE 3SAT } is
 $ \mathbf{NP} $-complete. Here, we ask the following question.

\begin{prob}
Is there any polynomial time reduction from {\it Cubic Monotone  NAE 3SAT} to {\it Cubic Bipartite  NAE Decomposition Problem}?
\end{prob}

Next, we focus on the applications of NAE and 1-in-Degree decompositions.\\
\underline{An application of the 1-in-Degree decomposition:}
A {\it zero-sum  vertex $k$-flow} is a vector in the null-space of the 0,1-adjacency matrix
of $G$ such that its entries belong to $\{\pm 1,\ldots,\pm (k-1)\}$.
It was shown that for a given
bipartite $(2,3)$-graph $G$, it is $ \mathbf{NP} $-complete to
decide whether the graph $G$ has a zero-sum vertex 3-flow \cite{MR3523349}.
We use from our hardness results on 1-in-Degree decompositions and prove that
for a given 3-regular bipartite graph $G$ determining whether $G$ has a zero-sum vertex $3$-flow is $ \mathbf{NP} $-complete.
\\ \\
\underline{An application of the NAE decomposition:}
For a given graph $G$, {\it The Minimum Edge Deletion Bipartition Problem} is to determine
 the minimum number of
edges of $G$ such that their removal leads to a bipartite
graph $H$. It was shown that this problem is
$ \mathbf{NP} $-hard even if all vertices have degrees 2 or 3 \cite{feder2011maximum}.
As an application of the NAE  decomposition, we show that if {\it Cubic Bipartite NAE Decomposition Problem }
   is $ \mathbf{NP} $-complete, then for a given 3-regular
 graph $G$, {\it The Minimum Edge Deletion Bipartition Problem} is
$ \mathbf{NP} $-hard.
\\ \\
$
  \text{\it Cubic Bipartite NAE Decomposition Problem }
                                        \leq^p_m    \text{\it   Cubic Monotone NAE 3SAT} \\
                                       \leq^p_m    \text{\it The Minimum Edge Deletion Bipartition Problem}.
$

Next, we generalize  the decompositions for vertex-weighted graphs.
A vertex-weighted graph  is a graph in which each vertex has been assigned a weight.
Let $G$ be a vertex-weighted graph and $w: V(G)\rightarrow \mathbb{Z}$ be its weight
function. A {\it 1-in-Degree coloring} for the graph $G$ is a function $f:V(G) \rightarrow \{0,1\}$ such
that for each vertex $v\in V(G)$, $\sum_{u\in N(v)}f(u)w(u)=1$. We show that 1-in-Degree coloring in vertex-weighted
graphs is harder than 1-in-Degree decomposition in simple graphs. For instance, although we show that if the given graph $G$ is a
bipartite graph  and does not have any  cycle of length congruent to 2 mod 4, then there is
a polynomial time algorithm to decide whether the graph $G$ has a 1-in-Degree decomposition, but for a
given vertex-weighted bipartite graph $G$, determining whether the graph $G$
has  a 1-in-Degree  coloring  is strongly  $ \mathbf{NP} $-complete, even if the given graph $G$  does
not have any  cycle of length congruent to 2 mod 4.

Finally, we study the edge version of the 1-in-Degree and the NAE decompositions.
An edge coloring $f:E(G)\rightarrow \{0,1\}$ for a graph $G$
is called {\it 1-in-Degree edge coloring} if and only if for every vertex $v$, $\sum_{e \ni v}f(e)=1$.
The graph $G$ has a  1-in-Degree edge coloring if and only if $G$ has a perfect matching.
Now, consider the edge version of the NAE. An edge coloring $f:E(G)\rightarrow \{red,blue\}$ of a graph $G$
is called {\it NAE edge coloring} if and only if for
every vertex $v$, there are edges $e$ and $e'$ incident
with $v$, such that $f(e)\neq f(e')$.
In this work, we show that for a given connected graph $G$ with $\delta(G)>1$, the graph $G$ has
a NAE edge coloring if and only if $G$ is not an odd cycle.

\subsection{The organization of the paper}

The remainder of this paper is organized as follows. Basic definitions and notation are provided
in Subsection \ref{Section2}. In Section \ref{Section3}, we introduce a new version of {\it Planar 1-in-3 SAT}
and we prove that this version is also $ \mathbf{NP} $-complete. In consequence of
this result, we show that for a given planar $(3,4)$-semiregular graph $G$ determining whether there is a
vector in the null space of the 0,1-incidence matrix of $G$ such that its entries belong
to $\{ \pm 1,\pm 2\}$ is $\mathbf{NP} $-complete.
Next, in Section \ref{Section4} we focus on the NAE and 1-in-Degree decompositions
of graphs.
In this section we introduce these decompositions.
Afterwards, we study their computational complexities. In the next section,  some applications of these decompositions
in computational complexity are presented. In subsection \ref{Section5.1}, we consider zero-sum vertex
flows and we prove that for a given bipartite 3-regular graph $G$ determining whether  $G$
has a zero-sum vertex $3$-flow is $\mathbf{NP} $-complete.
Also, In subsection \ref{Section5.2}, as another application of the 1-in-Degree decomposition, we
consider {\it The Minimum Edge Deletion Bipartition Problem} and prove its hardness.
Next, in Section \ref{Section6}, we consider the NAE and 1-in-Degree decompositions for weighted graphs and
finally in Section \ref{Section7}, the edge versions of the NAE and 1-in-Degree decompositions are studied.
The paper is concluded with some
remarks in Section \ref{Section8}.

\subsection{Notation} \label{Section2}

Throughout this paper all graphs are finite, simple and we follow \cite{MR1567289, MR1367739} for terminology and
notation  not defined here. Also, throughout the paper we denote $\{1,2,\cdots, k\}$ by $\mathbb{N}_k$.
We denote the vertex set and the edge set of
$G$ by $V(G)$ and $E(G)$, respectively. Also, we denote the maximum degree
and minimum degree of $G$ by $\Delta(G)$ and $\delta(G)$,
respectively.
Furthermore, for every $v\in V (G)$ and $X \subseteq V(G)$, $d(v)$, $N(v)$ and $N(X)$  denote the degree of $v$,  the neighbor set of $v$
and the set of vertices of $G$ which has a neighbor in $X$, respectively.
The {\it adjacency matrix} of a simple graph $G$ is an $n \times n$ matrix $A=[a_{ij}]$ where $n$ is the number of vertices, such that $a_{ij} = 1$ if the vertex $v_i$ is adjacent with the vertex $v_j$ and $0$ otherwise.
 The {\it incidence matrix} of a directed graph $G$ is an $n \times m$ matrix $B=[b_{ij}]$ where $n$ and $m$ are the number of vertices and edges respectively, such that $b_{ij} = -1$ if the edge $e_j$ leaves vertex $v_i$, $1$ if it enters vertex $v_i$ and $0$ otherwise. Similarly, for an undirected graph $G$, the incidence matrix is an $n \times m$ matrix $B=[b_{ij}]$
such that $b_{ij} = 1$ if the vertex $v_i$ and edge $e_j$ are incident and $0$ otherwise.
The  {\it null-space (kernel)} of a matrix $A$ is the set of all vectors $x$ for which $Ax = 0$.
An eigenvector of a square matrix $A$ is a non-zero vector $v$ that, when multiplied with $A$, yields a scalar multiple of $A$. The scalar multiplier is often denoted by $\lambda$ and we have $Av= \lambda v$. The number $\lambda$ is called the eigenvalue of $A$ corresponding to $v$.
An eigenspace  of a matrix $A$ is the set of all eigenvectors with the same eigenvalue, together with the zero vector.

A graph $G$ is a $(d, d + s)$-graph if the degree of every vertex of the graph
 $G$ lies in the interval $[d, d + s]$. A $(d, d + 1)$-graph is said to be semiregular.
A {\it total dominating set} of a graph  is a set of vertices such that all vertices in the graph
 have at least one neighbor in the dominating set. Also, a  {\it total perfect dominating set}
  of a graph $G = (V , E)$ is a subset
$S$ of $V(G)$ such that every vertex $v\in V(G)$ is adjacent to
exactly one vertex of $S$ (for every vertex $v$, $|N(v)\cap S|=1$). A {\it matching}
in a graph is a set of edges without common vertices and a {\it perfect matching}
is a matching which matches all vertices of the graph. A {\it trail} in a graph is a
walk without repeated edges and an {\it Eulerian cycle} is a  trail in a graph which visits every edge exactly once.
A graph $G = [X, Y ]$ is called {\it bipartite}, if the vertex set $V(G)$ can be
partitioned into two disjoint subsets $X$ and $Y$, i.e., $V(G) = X \cup Y \text{ and } X \cap Y =\emptyset $, such
that every edge in $E$ connects a vertex
from $X$ to a vertex from $Y$.

\section{The structure of the kernel of the incidence matrix} \label{Section3}

It is well-known that {\it Planar 1-in-3 SAT} is $ \mathbf{NP} $-complete, even
if we add the {\it backbone} between the variable vertices \cite{mulzer2008minimum}. In the
proof of the next theorem (Theorem \ref{T5}),  we introduce
a new version of {\it Planar 1-in-3 SAT} which has a "tree"  between the clause vertices and we prove that
this version is also $ \mathbf{NP} $-complete. In consequence of this result, we show that for a
given planar $(3,4)$-semiregular graph $G$ determining whether there is a
vector in the null-space of the 0,1-incidence matrix of $G$ such that its
entries belong to $\{ \pm 1,\pm 2\}$ is $\mathbf{NP} $-complete.

\begin{thm}\label{T5}
It is $ \mathbf{NP} $-complete to determine whether a given planar
$(3,4)$-semiregular graph has a zero-sum 3-flow.
\end{thm}

\begin{proof}{
Our proof consists of two steps. In the first step, we prove that
{\it   Monotone Planar Tree-like 1-in-3 SAT } is NP-complete, then in
the next step, we reduce this version to our problem in polynomial time.
\\
{{\bf Step 1.}}
We reduce  {\it Cubic Planar 1-in-3 SAT } into {\it   Monotone Planar Tree-like 1-in-3 SAT } in polynomial time.
Let $\Psi$ be an instance of {\it  Cubic Planar 1-in-3 SAT}.
Consider the following graph obtained from $\Psi$. The graph has one vertex for
each variable, one vertex for each clause and each clause vertex is connected by an edge to the  variable
vertices corresponding to the literals present in the clause. Note that this graph is planar.
Step by step, put a new edge between the clause vertices until if you add another
edge the graph becomes non-planar. Call the resultant graph $G$. Since every edge
in the graph $G$ is between two clause vertices or between a variable vertex and a clause vertex,
also, there is no edge between variable vertices, the induced subgraph on the set of
variable vertices is an independent set.
In other words the set of variable vertices with the set of edges that incident with
 them forms a galaxy (a galaxy or a star forest is a union of vertex disjoint stars).
So, between every two clause vertices there
is a path that consists of clause vertices.
Consequently, the induced subgraph on the set of clause vertices is connected (otherwise
we can add more edges between the clause vertices). Thus we can remove some edges from the graph
$G$ such that the induced subgraph on the set of clause vertices forms a tree. Call the
resultant planar graph $H_{\Psi}$. Note that in the next step, when we say that
the graph $H_{\Psi}$ is the graph which is obtained from the formula $\Psi$, we mean a graph which is obtained from
the
above-mentioned procedure form the formula $\Psi$.  This completes the first step.
\\
{{\bf Step 2.}}
In this step, we reduce  {\it Monotone Planar Tree-like 1-in-3 SAT} to our problem in polynomial time.
Let $\Psi$ be an instance of {\it  Monotone Planar Tree-like 1-in-3 SAT}
and $H_{\Psi}$ be the graph which is obtained from $\Psi$ (we explained how to obtain $H_{\Psi}$ from $\Psi$ in Step 1).
For every
clause $c$, $c\in C$, we denote the number of clause vertices which are adjacent
with the clause vertex $c$ in $H_{\Psi}$ by $\gamma(c)$. Also, for a vertex $v$, we say
that the zero-sum rule holds on the vertex $v$,
when the sum of assignments of all edges incident with the vertex $v$ is
zero. For a given formula $\Psi$, we construct a planar $(3, 4)$-semiregular graph $G_{\Psi}$ from $H_{\Psi}$, such that
the formula $\Psi$ has a 1-in-3 SAT satisfying assignment
if and only if the graph $G_{\Psi}$ has a zero-sum 3-flow.
For every variable $x$, $x\in X$, we create a cycle
 of length six with the vertices $v^x_1, w^x_1,v^x_2, w^x_2, v^x_{3}, w^x_{3}$, in that order.
Now, consider three copies $I(u^x_1), I(u^x_2), I(u^x_3)$ of the gadget shown in Figure \ref{I1} and connect each $u_i^x$ to $v_i^x$ for $i=1,2,3$. Call the resultant gadget $\mathcal{A}_x$. In the gadget $\mathcal{A}_x$ call the set of vertices of degree two, { free vertices}.

\begin{figure}[ht]
\begin{center}
\includegraphics[scale=.9]{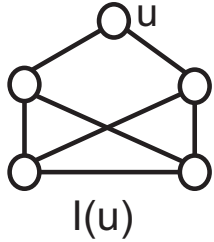}
\caption{The gadget $I(u)$.} \label{I1}
\end{center}
\end{figure}

For every clause $c$, $c\in C$, consider a
cycle of length $2\gamma(c)+8$  with the vertices $v^c_1, w^c_1,   \ldots, v^c_{\gamma(c)+4}, w^c_{\gamma(c)+4}$, in that order.
Now, consider $\gamma(c)+4$ copies $I(u^c_1), \ldots, I(u^c_{\gamma(c)+4})$ of the gadget
shown in Figure \ref{I1} and connect each vertex $u_i^c$ to the vertex $v_i^c$ for $i=1, \ldots, \gamma(c)+4$. Also, put a vertex $b_c$ and join it to the vertex $w^c_{\gamma(c)+4}$.
Call the resultant gadget $\mathcal{A}_c$. In the gadget $\mathcal{A}_c$ call the
set of vertices of degree two,  {\it free vertices} of $\mathcal{A}_c$
 and call the set of edges incident with the free vertices, {\it free edges}. For each
 pair $(g,h)$, where $g,h\in V(H_{\Psi})$ if $gh\in E(H_{\Psi})$, then join one of the free
 vertices of $\mathcal{A}_g$ to one of the free vertices of $\mathcal{A}_h$, such that having done these procedures
for all variables  and all clauses, the resultant graph is planar and its maximum
degree is three. After these procedures, for every clause $c\in C$, there are exactly
three free vertices of $\mathcal{A}_c $ that were joined to the free vertices of variables,
call these vertices, {\it important vertices} and join them to the vertex $b_c$. Note that
in the resultant graph the degree of the vertex $b_c$ is four and the graph is a planar $(3,4)$-semiregular graph. Finally,
for every {\it important vertex} $v$, choose one of the {\it free edges} incident
with the vertex  $v$ (note that every important vertex was a free vertex), suppose
that we choose $e=vz$, remove the edge $e$ from the graph and put a vertex $f_v$. Also, put
two copies $I(u_v), I(u_{v'})$ of the gadget shown in Figure \ref{I1} and connect
the vertex $f_v$ to the vertices $v, z, u_v, u_{v'}$. Call the resultant planar $(3,4)$-semiregular graph $G_{\Psi}$.

Now, suppose that the graph $G_{\Psi}$ has a zero-sum 3-flow. First, we present some useful lemmas.

\begin{lem}\label{lemma1}
For every vertex $t$ of degree three, the labels of the three edges incident with that vertex are $2,-1,-1$ or $-2,1,1$.
\end{lem}

{\bf Proof.}
For every vertex $t$ of degree three, the zero-sum rule
implies that not all three edges incident with that vertex can have odd labels. Since the label of each
edge is from $\{\pm 1, \pm 2\}$, the label $-2$ or the label 2 should
appear on exactly one of three edges incident the vertex $t$. Thus, by the zero-sum rule
the labels of the three edges incident with the vertex  $v$ are $2,-1,-1$ or $-2,1,1$.
$\spadesuit$
\\ \\
Similarly, for every vertex $t$ of degree four, the labels of the four edges
incident with that vertex are $2,-2,1,-1$ or $2,-2,2,-2$ or $1,-1,1,-1$. By above-mentioned
 lemma we have the following important lemma.

\begin{lem}\label{lemma4}
Let $t,t'\in V(G_{\Psi})$ be two vertices of degree three and $tt'\in E(G_{\Psi})$. Then the
labels of the three edges incident with the vertex $t$ are $2,-1,-1$ if and only if the labels
of the three edges incident with the vertex $t'$ are $2,-1,-1$.
\end{lem}

{\bf Proof.}
Since $t$ and $t'$ have a common edge, by Lemma \ref{lemma1}, the proof is clear.
$\spadesuit$

\begin{lem}\label{lemma2}
The set of edges with labels 2 or -2, in the induced subgraph on the set of vertices of degree three  forms a matching.
\end{lem}

{\bf Proof.}
By Lemma \ref{lemma1}, for every vertex $t$ of degree three, the labels of the three edges incident
with that vertex are $2,-1,-1$ or $-2,1,1$. So, the set of edges with labels 2 or $-2$, in the induced
subgraph on the set of vertices of degree three, forms a matching.
$\spadesuit$
\\ \\
Now, we study the main property of the gadget $I(u)$ which is shown in Figure \ref{I1}.

\begin{lem}\label{lemma3}
In each copy of the gadget $I(u)$ in the graph $G_{\Psi}$, the vertex $u$ has exactly one neighbor
other than its neighbors in $I(u)$. Without loss of generality call that vertex $t$. The
label of the edge $ut$ is 2 or $-2$.
\end{lem}

{\bf Proof.}
By Lemma \ref{lemma1} and Lemma \ref{lemma2}, the set of edges with labels 2 or -2, in the induced
subgraph on the set of vertices $V(I(u))\cup \{t\}$ forms a matching that saturates all the
vertices in $V(I(u))$. Since $I(u)$ has five vertices,
in every  matching $M$ for the induced subgraph on the set of vertices $V(I(u))\cup \{t\}$  that saturates all the
vertices in $V(I(u))$, we have $ut\in  M$.
Thus, the label of the edge $ut$ is 2 or $-2$.
$\spadesuit$

Let $ S\subseteq V(G_{\Psi})$ be a subset of vertices such that for every vertex $v\in S$, we have $d(v)=3$ and
 the induced subgraph on the set of vertices $S$ is connected. The zero-sum
rule implies that not all three edges incident with a vertex in $S$ can have odd labels, so -2 or
2 should appear on exactly one of them. Also, since the induced subgraph on the set
of vertices $S$ is connected,
without loss of generality we can suppose that every vertex in $S$ is incident with
exactly one edge with label 2 or every vertex in $S$ is incident with exactly one edge with
label $-2$. Therefore, for every $x\in X$, in the subgraph $\mathcal{A}_x$, every vertex
in $\mathcal{A}_x$ is incident with exactly one edge with label 2 or every vertex
in $\mathcal{A}_x$ is incident with exactly one edge with label $-2$.
Similarly, for each variable $x\in X$, by Lemma \ref{lemma3} and Lemma \ref{lemma4}
the set of labels of the edges between  $V(\mathcal{A}_x)$ and $V(G_{\Psi})\setminus V(\mathcal{A}_x)$ is $\{2\}$ or $\{-2\}$.
On the other hand,
since in the graph $H_{\Psi}$, the induced subgraph on the set of clause vertices forms a tree (is connected),
by Lemma \ref{lemma1} and Lemma \ref{lemma3}, in the graph $G_{\Psi}$, the set of labels of
edges $\{ w^c_{\gamma(c)+4} b_c| c\in C\}$ is $\{2\}$ or $\{-2\}$. Without loss of generality,
suppose that $\{ w^c_{\gamma(c)+4} b_c| c\in C\}=\{-2\}$ {\bf (Fact 1)}.

See Figure \ref{I2}. In this subgraph by Lemma \ref{lemma3} and lemma \ref{lemma4}, The labels of colored edges are $\pm 2$ and the labels of  black edges are $\pm 1$.

\begin{figure}[ht]
\begin{center}
\includegraphics[scale=.5]{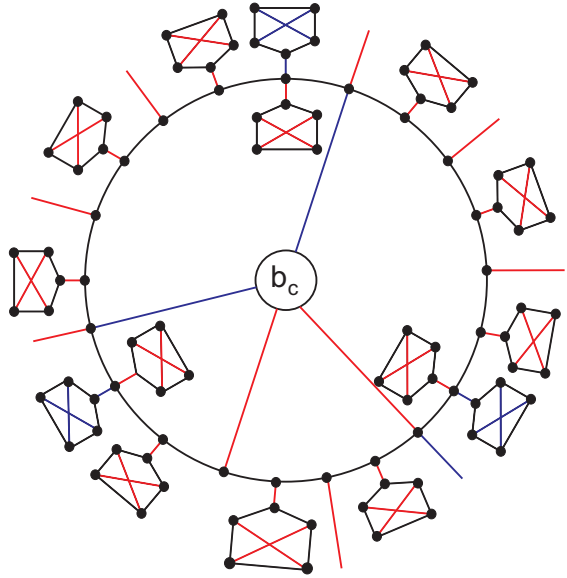}
\caption{The labels of blue edges are 2, the labels of red edges are $-2$ and the labels of  black edges are $\pm 1$.
Note that for every important vertex $v$, because of the vertex $f_v$, the labels of the four edges
incident with the vertex $v$ are $2,-2,1,-1$.} \label{I2}
\end{center}
\end{figure}

Now, we present a 1-in-3 SAT satisfying assignment for $\Psi$.
For every $x\in X$,
if every vertex in $\mathcal{A}_x$ is incident with exactly one edge
with label 2, put $\Gamma(x)={\sf true}$ and if every vertex in $\mathcal{A}_x$ is
incident with exactly one edge with label $-2$, put $\Gamma(x)={\sf false}$. For every
clause $c\in C$, the vertex $b_c$ is incident with the edges with
labels from $\{-2,2\}$. By Fact 1, the label of edge $ w^c_{\gamma(c)+4} b_c$ is $-2$, thus
the zero-sum rule implies that  the labels of edges incident with $b_c$, other
than $w^c_{\gamma(c)+4} b_c $ are exactly $2,2,-2$. Thus $\Gamma$ is a 1-in-3 SAT satisfying assignment for $\Psi$.
Conversely, if $\Psi$
has a 1-in-3 SAT satisfying assignment $\Gamma$, we may assign
labels $-2$ to the set of  edges $\{ w^c_{\gamma(c)+4} b_c| c\in C\}$.
Next, for every variable $x$, if $\Gamma(x)={\sf true}$ (respectively, ${\sf false}$), label the edges of $\mathcal{A}_x$ such that each vertex in $\mathcal{A}_x$ is incident with exactly one edge with label $2$ (respectively, $-2$).  It is easy to extent this labeling to a zero-sum 3-flow. This completes the proof.

}\end{proof}


\section{Not-All-Equal and 1-in-Degree decompositions}\label{Section4}

In the next theorem, we consider the computational complexity of the determining whether a graph has a
 1-in-Degree decomposition.

\begin{thm}\label{T1}\\
$(i)$ For every $r\geq 3$, for a given $r$-regular bipartite graph $G$
which  $G$ has a   Not-All-Equal decomposition determining
whether $G$ has a 1-in-Degree decomposition is  $ \mathbf{NP} $-complete.\\
$(ii)$ If $G$ is a bipartite graph   and does not have any  cycle of length
congruent to 2 mod 4, then there is a polynomial time algorithm to decide
whether the graph $G$ has a 1-in-Degree decomposition.
\end{thm}

\begin{proof}{

$(i)$
The problem is in $ \mathbf{NP} $. We reduce {  Cubic Planar 1-in-3 SAT} to our problem.
 Moore and Robson \cite{MR1863810} proved
that the following problem is $ \mathbf{NP}$-complete.
\\ \\
 {\it Cubic Planar 1-in-3 SAT.}\\
\textsc{Instance}: A 3SAT formula $\Phi=(X,C)$
 such that every variable
 appears in exactly three clauses, there
 is no negation in the formula, and the
bipartite graph obtained by linking a variable and a clause if and only
 if the
 variable appears in the clause, is planar.\\
\textsc{Question}: Is there a truth assignment for $X$ such that
 each clause in $C$ has exactly
one {\sf true} literal?
\\ \\
On the other hand, Moret   \cite{p} proved that  {\it Planar NAE 3SAT}
is in $ \mathbf{P} $ by an interesting reduction to a known problem in $ \mathbf{P} $, namely Planar MaxCut.
We  also use Moret's result in our proof.
\\ \\
 {\it Planar NAE 3SAT.}\\
\textsc{Instance}: A 3SAT formula $(X,C)$  such that the following graph obtained from 3SAT is planar. The graph has one vertex for each variable, one vertex for each clause; all variable vertices are connected
in a simple cycle and each clause vertex is connected by an edge to variable
vertices corresponding to the literals present in the clause (note that positive
and negative literals are treated exactly alike).\\
\textsc{Question}: Is there a NAE truth assignment for $X$?
\\ \\
By a simple argument we can see  that the reduction holds also for the following problem.
So the following problem is in  $ \mathbf{P} $ (for more information see \cite{ziad, Dehghan2015}).
\\ \\
{\it Planar NAE 3SAT Type 2.}\\
\textsc{Instance}: A 3SAT formula $(X,C)$ such that  the following graph obtained from 3SAT is planar.
The graph has one vertex for each variable, one vertex for each clause and each clause vertex is connected by an edge to variable
vertices corresponding to the literals present in the clause.\\
\textsc{Question}: Is there a NAE truth assignment for $X$?\\

Let $\Phi$ be an instance of {\it  Cubic Planar 1-in-3 SAT}. We can check in polynomial
time whether $\Phi$ has a NAE SAT satisfying  assignment. If $\Phi$ does not have
any NAE SAT satisfying assignment, then it does not have any 1-in-3 SAT satisfying  assignment.
So, suppose that $\Phi$ has a NAE SAT satisfying assignment. Let $r\geq 3$ be a fixed number. For
a given formula $\Phi$ we construct an $r$-regular bipartite graph $G$ such that the formula  $G$ has
a  1-in-Degree decomposition if and only if the formula  $\Phi$ has a 1-in-3 SAT satisfying assignment.
Our proof consists of two steps.\\
{{\bf Step 1.}}
Without loss of generality suppose that the number of clauses in $\Phi$ is $s$.
For each $j$, $j\in \mathbb{N}_{(r-3)s}$, consider the following $r$ clauses:
\\
\\
$\circ$ For every $i$, $i\in \mathbb{N}_{r- 1}$, consider the clause $(\varepsilon_{i}^{j} \vee \bigvee_{k=1}^{r-1} \alpha_{k}^j  )$.  \\
$\circ$  The clause $(\alpha_{1}^{j} \vee \bigvee_{k=1}^{r-1} \varepsilon_{k}^j  )$.
\\
\\
Assume that the set of above-mentioned clauses has a 1-in-$r$ SAT satisfying assignment $\Gamma $.
If there is an index $i$ such that $\Gamma(\varepsilon_{i}^{j})={\sf true}$, then
$\Gamma( \alpha_1^j)=\Gamma( \alpha_2^j)=\cdots=\Gamma( \alpha_{r-1}^j)={\sf false}$. So
$\Gamma(  \varepsilon_1^j)=\cdots= \Gamma(\varepsilon_{r-1}^j) ={\sf true}$.
Thus, $\Gamma$ is not a 1-in-$r$ SAT satisfying assignment for the
clause $(\alpha_{1}^{j} \vee \bigvee_{k=1}^{r-1} \varepsilon_{k}^j  )$.
This is a contradiction.
Similarly, if there is an index $i>1$ such that $\Gamma( \alpha_i^j)={\sf true}$, then
$\Gamma$ is not a 1-in-$r$ SAT satisfying assignment for the
clause $(\alpha_{1}^{j} \vee \bigvee_{k=1}^{r-1} \varepsilon_{k}^j  )$.
Hence, we have
$\Gamma( \alpha_2^j)=\Gamma( \alpha_3^j)=\cdots=\Gamma( \alpha_{r-1}^j)=
\Gamma(  \varepsilon_1^j)=\cdots= \Gamma(\varepsilon_{r-1}^j) ={\sf false}$. Call
the  set of above  clauses $\Upsilon$ and their variables $X'$.
Now, consider $r(r-2)$ copies of $\Phi$ and a copy of $\Upsilon$. For each
clause $c$ with $3$ variables,  add $r-3$  variables  from $ X'$ to $c$, such
that in the resultant formula no variable  appears in more then $r$ clauses. Call
 the resultant formula $\Phi'$. Each clause in $\Phi'$ has $r$ variables and every variable appears in exactly $r$ clauses.
Since for each $j$, $\Gamma( \alpha_2^j)=\Gamma( \alpha_3^j)=\cdots=\Gamma( \alpha_{r-1}^j)=
\Gamma(  \varepsilon_1^j)=\cdots= \Gamma(\varepsilon_{r-1}^j) ={\sf false}$, the formula  $\Phi'$ has a 1-in-$r$ SAT satisfying
 assignment if and only if the formula $\Phi $ has a 1-in-3 SAT satisfying assignment. This completes the first step.
\\
{{\bf Step 2.}}
Let $\Phi'$ be a given formula with the set of variables $X$ and the set of clauses $C$. Define the following graph.\\

$\begin{aligned}
V(G') &=\{\beta^{k}_{i,j,x},\,\delta^{k}_{1,j,x},\, x^{k}_j| 0\leq i \leq r-2,\, 0\leq j,k \leq r-1,\, x\in X\}\\
       &\cup \{\alpha^{k}_{i,j,x},\, c^{k},\varepsilon_{j,x}^{k},\zeta_{j,x}|0\leq i \leq r-3,\,
        0\leq j,k \leq r-1,\, c\in C,\, x\in X\}.
\end{aligned}$

$\begin{aligned}
E(G')&=\{x^{k}_j \alpha^{k}_{i,j,x}|  0\leq i \leq r-3,\, 0\leq j,k \leq r-1,\, x\in X \}\\
     &\cup \{ \alpha^{k}_{i,j,x} \beta^{k}_{i',j,x} ,\,  \beta^{k}_{i',j,x}\delta^{k}_{1,j,x} | 0\leq i \leq r-3,\, 0\leq i'
 \leq r-2,\, 0\leq j,k \leq r-1,\, x\in X\}\\
     &\cup \{\delta^{k}_{1,j,x} x^{k}_{(j+1 \mod r)} |  0\leq j,k \leq r-1,\, x\in X \}\\
     &\cup \{\varepsilon_{j,x}^k \beta^{k}_{i,j,x}| 0\leq i \leq r-2,\, 0\leq j,k \leq r-1,\, x\in X \}\\
     & \cup \{\varepsilon_{j,x}^{k}\zeta_{j,x}| 0\leq j,k \leq r-1,\, c\in C,\, x\in X \}.
\end{aligned}$

Now, we discuss the basic properties of the graph $G'$. Remove the set of
vertices $\{\zeta_{j,x}| 0\leq j \leq r-1,\, x\in X\}$ from the graph $G'$ and call the resultant graph $G''$.
The vertices of the graph $G''$ can be partitioned into $r$ parts such that there is no edge between two parts
and each part is a copy of the graph $J(z)$ which is shown in \ref{Fig2000} (it is enough to partition the vertices based on
parameter $k$). In the graph $J(z)$, $z$ is a fixed number. We have the following lemma about the graph $J(z)$.

\begin{lem}\label{lemma5}
For every copy of the gadget $J(z)$, define the following function $g:V(J(z))\rightarrow \{{\sf true},{\sf false}\}$ for
the vertices of $J(z)$
such that
$g(v)={\sf true}$ if the vertex $v$ is a white vertex,  $g(v)={\sf false}$ if the vertex $v$ is a black vertex and for
each two vertices $v$ and
$u$, if $v $ is a red vertex and $u $ is a blue vertex, then $g(v)\neq g(u)$.
In the  function $g$ each vertex of degree $r$ has exactly one neighbor  with label ${\sf true}$.
\end{lem}

{\bf Proof.}
In the graph $J(z)$, each vertex of degree $r$ has exactly one white neighbor or a blue neighbor and a red neighbor.
Thus, in the  function $g$ each vertex of degree $r$ has exactly one neighbor  with label ${\sf true}$.
$\spadesuit$

\begin{figure}[ht]
\begin{center}
\includegraphics[scale=.4]{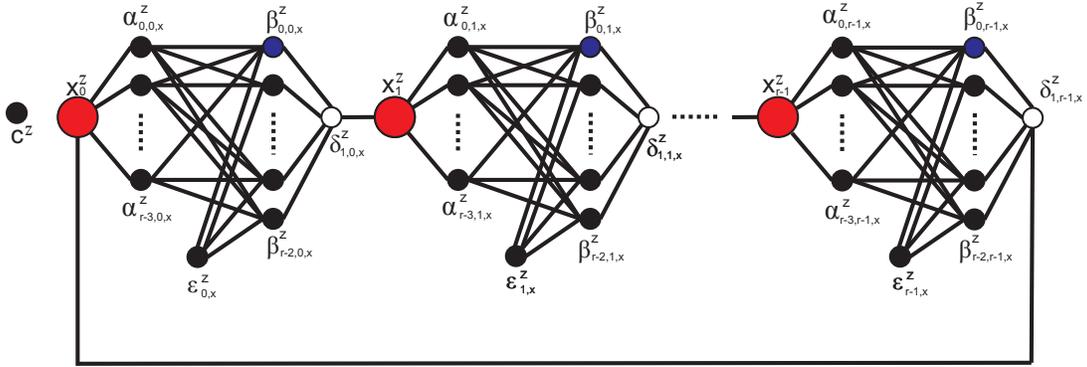}
\caption{The bipartite graph $J(z)$.} \label{Fig2000}
\end{center}
\end{figure}

Assume that the function $g$ in each copy  of the gadget $J(z)$, labels the set of red vertices with ${\sf true}$.
Define $\mathcal{S}_{1}=\{\varepsilon_{j,x}^{r-1},\,  c^{r-1},\zeta_{j,x}|
        0\leq j \leq r-1,\, c\in C,\, x\in X\}$ and $\mathcal{S}_{2}=\{\delta^{r-1}_{1,j,x}|
        0\leq j \leq r-1,\, x\in X\}$.
Now, consider the graph $G'$ and using $\mathcal{S}_{1},\mathcal{S}_{2}$ define the function $h:V(G')\rightarrow \{{\sf true},{\sf false}\}$:\\
$$
h(v)=\begin{cases}
{\sf true},      &{\text{     if  }}\,\,  v\in \mathcal{S}_{1}, \\
{\sf false},     &{\text{     if  }}\,\,  v\in \mathcal{S}_{2}, \\
g(v),      &{\text{        }}\,\,  otherwise.
\end{cases}
$$

In the graph $G'$ the degree of each vertex except the set of
vertices $\{x^{k}_j,c^{k}| 0\leq j,k \leq r-1,\, c\in C,\, x\in X\}$ is $r$.
We have the following lemma for the function $h$.

\begin{lem}\label{lemma6}
In the function
$h$ each vertex of degree $r$ in the graph $G'$ has exactly one neighbor  with label ${\sf true}$.
\end{lem}

{\bf Proof.}
Since in the function $g$ each vertex of degree $r$ in $G''$ has exactly one neighbor  with label ${\sf true}$, in the function
$h$ each vertex of degree $r$ in the graph $G'$ has exactly one neighbor  with label ${\sf true}$.
$\spadesuit$

See Figure \ref{I5}, to see the structure of bipartite graph $G'$, where $r=3$, $C=\{c\}$ and $X=\{x\}$.

\begin{figure}[ht]
\begin{center}
\includegraphics[scale=.4]{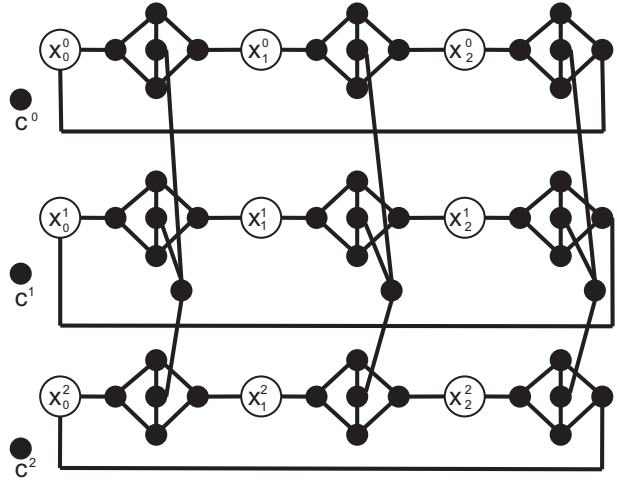}
\caption{The bipartite graph $G'$ for $r=3$, $C=\{c\}$ and $X=\{x\}$.} \label{I5}
\end{center}
\end{figure}

Now, consider a copy of the graph $G'$. In the graph $G'$ the degree of each vertex except the set of
vertices $\{x^{k}_j,c^{k}| 0\leq j,k \leq r-1,\, c\in C,\, x\in X\}$ is $r$.
For every clause $c$ and variable $x$ if the variable $x$ appears in the clause $c$, then for each $k$, $0\leq k \leq r-1 $,
join the vertex  $c^k$ to exactly one of the vertices $x^{k}_j$ such that in the resultant graph the
degree of each vertex is $r$. Call the resultant graph $G$.
The graph $G$ is an $r$-regular bipartite graph.
First, assume that  $\Gamma : X \rightarrow \{ {\sf true},{\sf false}\} $ is a 1-in-$r$  SAT satisfying assignment for $\Phi'$.
Now, we present a 1-in-Degree decomposition for the graph $G $.
Define the function $f$ as follows.

$f(x_j^k)={\sf true}$, if and only if $\Gamma(x)={\sf true}$,

$f(\alpha^{k}_{i,j,x} )={\sf false}$,

$f(\beta^{k}_{i,j,x} )={\sf false}$, for $i\neq 0$,

$f(\beta^{k}_{0,j,x} )={\sf false}$,  if and only if $\Gamma(x)={\sf true}$,

$f(\varepsilon_{j,x}^{k})={\sf false}$ and  $f(\delta^{k}_{1,j,x})={\sf true}$, $f(c^k)={\sf false}$, for $k\neq  r-1$,

$f(\varepsilon_{j,x}^{r-1})={\sf true}$  and $f(\delta^{r-1}_{1,j,x})={\sf false}$, $f(c^{r-1})={\sf true}$,

$f(\zeta_{j,x})={\sf true}$,   if and only if $\Gamma(x)={\sf true}$.
\\ \\
The function $f$ is similar to the function $h$ except that\\
$\bullet $ For each $(j,k)$,  $f(x_j^k)={\sf true}$ and $f(\beta^{k}_{0,j,x} )={\sf false}$  if and only if $\Gamma(x)={\sf true}$.\\
$\bullet $ For each $(j,x)$, $f(\zeta_{j,x})={\sf true}$, if and only if $\Gamma(x)={\sf true}$.

Thus, by Lemma \ref{lemma6}, the function $f$ is a  1-in-Degree decomposition for the graph $G $
(note that the graph $G $ always has a  NAE decomposition).

Next, suppose that the graph $G $ has  a 1-in-Degree decomposition $f:V(G)\rightarrow \{{\sf true},{\sf false}\}$. For   every
variable $x\in X$,  we have the following important lemma.

\begin{lem}\label{lemma7}
For every
variable $x\in X$, $f(x_0^1)=f(x_1^1)=\cdots=f(x_{r-1}^1)$.
\end{lem}

{\bf Proof.}
In the graph $G $,  the red vertices are connected through   some paths of length four.
Let $ k_1$ and $k_2$ be two red vertices and $k_1 s_1 s_2 s_3k_2$ be a path of length four.
First, assume that the label of $k_1$ is ${\sf true}$. By this assumption, the labels of all
the neighbors of the vertex $s_1$ except the vertex $k_1$ are ${\sf false}$.
Since $N(s_3)=\{k_2\}\cup N(s_1)\setminus \{k_1\} $, the label of the vertex $k_2$ is ${\sf true}$. Similarly, if
the label of the vertex  $k_1$ is ${\sf false}$, one can see that the label of the vertex  $k_2$ is ${\sf false}$.
$\spadesuit$

Now we present a 1-in-$r$  SAT satisfying assignment for $\Phi'$
Define the function $\Gamma : X \rightarrow \{{\sf true},{\sf false} \}$ such that for each variable $x$
$\Gamma(x)={\sf true}$ if and only if $f(x_0^1)={\sf true}$.
By Lemma \ref{lemma7} and since $f$ is a symmetric 1-in-$d$ SAT satisfying
assignment, $\Gamma$ is a 1-in-$r$  SAT satisfying assignment for $\Phi'$.
This completes the proof.
\\
\\
$(ii)$ Let $G $ be a given graph. Since the graph
$G $ is a bipartite graph, the following integer linear program
determines whether the graph $ G$ has a 1-in-Degree decomposition:

\begin{equation*}
\begin{array}{lll@{}ll}
\text{minimize}  &                    1    & &  &\\
\text{subject to}& & \sum_{u \in N(v)} \, \,f(u)   &\geq 1,             &\forall v\in V(G)\\
                 &  &\sum_{u \in N(v)}-f(u)  &\geq -1,            &\forall v\in V(G)\\
                 &  & \, \,\, \,\, \,\, \,\, \,\, \,\, \,\, \,\, \,\, \, \, \,\, \, f(v)                &  \in \{0,1\},      &\forall v\in V(G)
\end{array}
\end{equation*}

The above integer linear program is feasible if and only if the graph $G$ has a 1-in-Degree decomposition.
When an integer linear program has all-integer coefficients and the matrix
of coefficients is totally unimodular, then the optimal solution of its relaxation is
integral \cite{book2}. Hence, it can be obtained in polynomial time.
A { signed bipartite graph} is a bipartite graph in which every edge is given a label of $-1$ or
$+1$. The weight of a cycle in a signed bipartite graph is the sum of the labels of its edges. A
signed bipartite graph is called { restricted unimodular} if the weight of any cycle in the graph  $G$ is divisible
by 4. There is a  correspondence between a signed bipartite graph $G$ and its $\{-1, 0, 1\}$
adjacency matrix $D$, where the rows and columns of $D$ are indexed by the vertices of the graph  $G$, with
$D_{ij} = -1, 0, 1$, according as the edge between vertices $v_i$ and $v_j$ has label $-1$, is absent, or has
label $+1$, respectively.  Commoner \cite{TU1} proved that
a signed bipartite graph that is restricted unimodular has a totally unimodular
adjacency matrix. If  the graph $G$ is a bipartite graph,
 does not have any  cycle of length congruent to 2 mod 4 and each edge  label is 1, then
 the graph $G_{\Phi}$ is restricted unimodular and  has a totally unimodular
adjacency matrix. Consequently,
there is polynomial time algorithm to determine whether the above-mentioned integer linear program is feasible.
This completes the proof.

}\end{proof}

A {\it Hypergraph} is a pair $\mathcal{H} = (V,E)$ such that $E$ is a
subset of the power set of $V$. The set $V$ is the set of vertices and $E$ is the set
of edges. A proper $l$-coloring of a hypergraph $\mathcal{H}$ is a function $c : V(\mathcal{H})\rightarrow \mathbb{N}_{l}$ in which there is no monochromatic
edge in $\mathcal{H}$. We say that a hypergraph $\mathcal{H}$ is $t$-colorable if there is a proper $t$-coloring of it.
If $E$ contains only sets of size $k$ then $\mathcal{H}$ is said to be {\it $k$-uniform} and if every vertex appears in exactly $r$
edges then $\mathcal{H}$ is said to be {\it $r$-regular}.
The main result by Thomassen \cite{MR1135027} implies
that every $r$-uniform, $r$-regular hypergraph is $2$-colorable for all $r \geq 4$ (for more information see \cite{MR965385}).
Therefore, we have the following result:

\begin{pro}\label{T2} \cite{MR1135027}
If $G $ is an $r$-regular bipartite graph and $r\geq 4$, then  the graph $G$ has a NAE decomposition.
\end{pro}

There are $3$-uniform $3$-regular hypergraphs that are not $2$-colorable. For
instance, consider the Fano Plane. The Fano Plane is a hypergraph  with seven vertices $ \mathbb{Z}_{7} $
and seven edges $ \{ \{ i,i+1,i+3 \}:   i\in \mathbb{Z}_{7} \}$. The Fano Plane is not $2$-colorable  and is minimal with respect
to this property  \cite{bondy}.

\begin{remark}{
There is another interesting problem which is related to the NAE decomposition.
A vertex-labeling $f$ is a {\it gap labeling} if

$c(v)=\begin{cases}
                   1       & $if$ \,\,d(v)=0,\\
    f(u)_{u  v \in E(G)}       & $if$ \,\,d(v)=1,\\
    \max_{u  v \in E(G)} f(u) - \min_{u  v \in E(G)} f(u)       & $otherwise$,\
\end{cases}$
\\
is a proper vertex coloring \cite{MR3072733}. Every bipartite graph $G=[X,Y]$ has
a {\it gap labeling}, label the set of vertices $X$ by one
and label the set of vertices   $Y$ by different even numbers. In \cite{MR3072733}
it was asked, to determine the computational complexity of deciding whether a given 3-regular bipartite graph $G$ have a
{\it gap labeling} from $ \mathbb{N}_{2}$. In other words, given a
formula $\Phi$ in conjunctive normal form, such that each clause contains three
literals and each literal appears in exactly three clauses and the formula is monotone,
determine the computational complexity of deciding whether there exists a truth assignment
to the variables and clauses so that  each clause has at least
one {  true} literal and at least one {  false} literal or each variable  appears in at
least one {  true} clause and at least one {  false} clause.
}\end{remark}

\begin{thm}\label{T3}
If $G $ does not have any  cycle of length congruent to 2 mod 4, then
there is a polynomial time algorithm to decide whether $G$ has a
 NAE decomposition.
\end{thm}

\begin{proof}{

Let $G $ be a graph. The following integer linear
program determines whether $G$ has a  NAE decomposition:

\begin{equation*}
\begin{array}{lll@{}ll}
\text{minimize}  &                    1    & &  &\\
\text{subject to}& & \sum_{u \in N(v)}\, \,f(u)   &\geq 1,             &\forall v\in V(G)\\
                 &  &\sum_{u \in N(v)}-f(u)  &\geq -d(v)+1,       &\forall v\in V(G)\\
                 &  &   \, \,\, \,\, \,\, \,\, \,\, \,\, \,\, \,\, \,\, \, \, \,\, \,f(v)                &  \in \{0,1\},      &\forall v\in V(G)
\end{array}
\end{equation*}

The above-mentioned integer linear program is feasible if and only if $G$ has
a NAE decomposition. The other parts of the proof are
similar to the proof of Theorem \ref{T1}, Part $(ii)$.

}\end{proof}

\section{Applications of the 1-in-Degree and NAE decompositions}\label{Section5}

\subsection{The structure of the kernel of the adjacency matrix} \label{Section5.1}

A {\it zero-sum vertex flow} of a graph $G$ is an assignment of non-zero integer
numbers to the vertices of $G$ such that  the sum of the labels of all vertices adjacent with
each vertex is zero.
Let $k$ be a natural number. A {\it zero-sum  vertex $k$-flow} is a zero-sum vertex flow with
labels from the set $\{\pm 1,\ldots,\pm (k-1)\}$.
The first natural problem about the computational complexity of the existence of
zero-sum  vertex  flow in graphs is the following problem.

\begin{prob} \cite{MR3523349}
Is there a polynomial time algorithm to decide whether a given graph $G$ has  a zero-sum vertex flow?
\end{prob}

It was shown that for a given
bipartite $(2,3)$-graph $G$, it is $ \mathbf{NP} $-complete to
decide whether the graph $G$ has a zero-sum vertex 3-flow \cite{MR3523349}.
Here, we improve the previous complexity result.
A given 3-regular bipartite graph $G$  has a zero-sum vertex $3$-flow if and only if  the graph $G$
has a  1-in-Degree decomposition.
Therefore, by Theorem \ref{T1}, we have the following:

\begin{cor}
For a given 3-regular bipartite graph $G$ determining whether $G$ has a zero-sum vertex $3$-flow is $ \mathbf{NP} $-complete.
\end{cor}

\subsection{The Minimum Edge Deletion Bipartition Problem} \label{Section5.2}

For a given graph $G$, {\it The Minimum Edge Deletion Bipartition Problem} is to determine
 the minimum number of
edges of $G$ such that their removal leads to a bipartite
graph $H$. It was shown that the minimum edge deletion bipartition problem is
$ \mathbf{NP} $-hard even if all vertices have degrees 2 or 3 \cite{feder2011maximum}.
Here, we show the following:

\begin{thm}\label{T11}
If {\it Cubic Bipartite NAE Decomposition Problem }   is $ \mathbf{NP} $-complete, then for a given 3-regular
 graph $G$, {\it The Minimum Edge Deletion Bipartition Problem} is
$ \mathbf{NP} $-hard.
\end{thm}

\begin{proof}{

Suppose that {\it Cubic Bipartite NAE Decomposition Problem } is
 $ \mathbf{NP} $-complete. There is a simple polynomial time
 reduction from {\it Cubic Bipartite NAE Decomposition Problem } to {\it Cubic Monotone  NAE 3SAT}, thus,
 {\it Cubic Monotone  NAE 3SAT }
 is $ \mathbf{NP} $-complete. We reduce {\it Cubic Monotone  NAE 3SAT} to our problem in polynomial time. The proof, given below, follows
the same approach as \cite{feder2011maximum}.

Let $\Phi$ be a given formula with the set of variables $X$ and the set of clauses $C$. Without loss of generality,
 suppose that $|C|=k$. We construct a graph $G$ such that  the minimum number of
edges of $G$ such that their elimination leads to a bipartite
graph is exactly $k$ if and only if the formula $\Phi$ has a  NAE SAT satisfying assignment.
For every variable $x$, put a vertex $r_x$ in the graph. Also, for every
pair $(x,c)$, where $x\in X$ and $c\in C$, if $x$ appears in $c$ then put
a vertex $c_x$ in the graph and join the vertex $c_x$ to the vertex $r_x$.
Finally, for every clause $c=(x \vee y \vee z)$, join the vertex $c_x$ to
the vertices $c_y$ and $c_z$, also, join the vertex $c_y$ to the vertex $c_z$. Call
the resultant 3-regular graph $G$. Since the graph $G$ contains at least $k$ edge disjoint triangles,  the minimum number of
edges of the graph $G$ such that their removal leads to a bipartite
graph is at least $k$.
First, suppose that the instance $\Phi$ has a
NAE assignment $\Gamma$. For each clause $c=(x \vee y \vee z)$, the triple $(x,y,z)$ in the formula $\Gamma$ has
either two ${\sf true}$s and a ${\sf false}$, or two ${\sf false}$s and a ${\sf true}$. From the three edges $c_xc_y,c_yc_z, c_zc_x$,  remove the
edge joining the two ${\sf true}$s or
the two ${\sf false}$s.
Now, we show that the remaining graph is bipartite. Consider the following partition for the vertices of the graph. For every
variable $x$, put the vertex $r_x$ in $ F$ if and only if $\Gamma(x)={\sf true}$. Also, for every vertex $c_x$ put the vertex
$c_x$ in $F$ if and only if $\Gamma(x)=flase$. One can see that $(F, \overline{F})$ is a partitioning for the vertices of the graph
such
that  every edge in the remaining graph connects a vertex
from $F$ to a vertex from $\overline{F}$.
Thus, the remaining graph is bipartite. Next, assume
that $G$ has a solution that removes exactly one edge per triangle to obtain a bipartite
graph $H$. Label the two sides of the bipartition of $H$ by $1$
and $-1$ respectively. Call this labeling $\ell$. For every pair $(x,c)$, where
$x\in X$ and $c\in C$, if $x$ appears in $c$, by the structure of $H$, we
have $\ell(c_x)=-\ell(r_x)$. For every variable $x$, put $\Gamma(x)={\sf true}$ if and
 only if $\ell(r_x)=-1$. One can see that the function  $\Gamma$ is a NAE SAT satisfying assignment for the formula $\Phi$.

}\end{proof}

\section{1-in-Degree decomposition in vertex-weighted graphs}\label{Section6}

A vertex-weighted graph  is a graph in which each vertex has been assigned a weight.
Let $G$ be a vertex-weighted graph and $w: V(G)\rightarrow \mathbb{Z}$ be its weight
function. A {\it 1-in-Degree coloring} for the graph $G$ is a function $f:V(G) \rightarrow \{0,1\}$ such
that for each vertex $v\in V(G)$, $\sum_{u\in N(v)}f(u)w(u)=1$. Although, if the graph $G$ is a
bipartite graph  and does not have any  cycle of length congruent to 2 mod 4, then there is
a polynomial time algorithm to decide whether the graph $G$ has a 1-in-Degree decomposition, but for a
given vertex-weighted bipartite graph $G$, determining whether the graph $G$
has  a 1-in-Degree  coloring  is strongly  $ \mathbf{NP} $-complete, even if the graph $G$  does
not have any  cycle of length congruent to 2 mod 4.

\begin{thm}\label{T7}
For a given vertex-weighted bipartite graph $G$, determining
whether $G$ has  a 1-in-Degree   coloring  is strongly  $ \mathbf{NP} $-complete, even
if the graph $G$  does not have any  cycle of length congruent to 2 mod 4.
\end{thm}

\begin{proof}{
It was shown that {\it 3-Partition} is $ \mathbf{NP}$-complete in the strong sense \cite{MR1567289}.
\\ \\
{\it 3-Partition.}\\
\textsc{Instance}: A positive integer $ k \in \mathbb{Z}^{+}$ and $3n$ positive
 integers $a_1, \ldots, a_{3n} \in \mathbb{Z}^{+}$ such that for each $i$, $1 \leq i \leq 3n$,  $  k/4 < a_i < k/2$. Also $ \sum _{i=1}^{3n} a_i=nk$.\\
\textsc{Question}: Can $\{a_1, \ldots, a_{3n}\}$ be partitioned into $n$ disjoint sets $A_1, \ldots , A_n$ such that
for each $i$, $i \in \mathbb{N}_{n}$, $ \sum _{a \in A_i}a= k$?

We reduce {\it 3-Partition} to
our problem in polynomial time.
For an instance $A= [ a_1, \ldots,a_{3n} ]$ and number $k$, define the
bipartite graph $G$ with the weight function $w$ as follow:
\\
\\
$V(G)=\{x_{i,j}, y^t_i, z^{l}_j|   i \in \mathbb{N}_{ n}, \,   j \in \mathbb{N}_{3n},\,   t \in \mathbb{N}_{2} ,\,
  l \in \mathbb{N}_{5}\}$.
\\
\\
$E(G)=\{y^1_i x_{i,j},\, y^1_i y^2_i,\, z^1_j x_{i,j},\, z^1_j z^2_j, z^2_j z^3_j,
z^2_j z^4_j,z^3_j z^5_j, z^4_j z^5_j |  i \in \mathbb{N}_{ n}, \,   j \in \mathbb{N}_{3n}\}$.
\\
\\
$w(x_{i,j})=a_j, \,\, w(y^1_i)=1,\,\, w(y^2_i)=1-k, \, \,w(z^1_j)=w(z^3_j)=w(z^4_j)=1, \,\, w(z^2_j)=1-a_j,\,\, w(z^5_j)=a_j$.
\\
\\
The graph $G$  does not have any  cycle of length congruent to 2 mod 4. See Figure \ref{NNF2}

\begin{figure}[ht]
\begin{center}
\includegraphics[scale=.6]{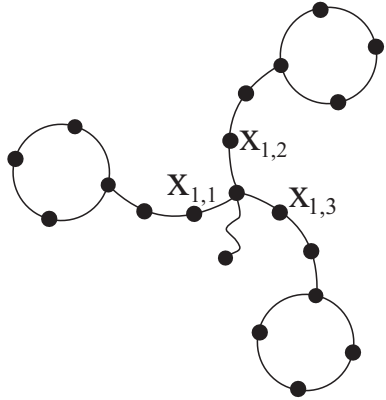}
\caption{The graph $G$ for $n=1$.}\label{NNF2}
\end{center}
\end{figure}

Let $f:V(G)\rightarrow \{1,0\}$ be a 1-in-Degree coloring for the graph $G$. For each $j$, $  j  \in \mathbb{N}_{3n}$ we have:

\begin{align*}
1                                         &=\displaystyle \sum_{u\in N(z^1_j)}f(u)w(u) \\
                                          &= (1-a_j)f(z^2_j)+(a_j)\sum_{i}f(x_{i,j})\\
                                          &= \sum_{i}f(x_{i,j})      & {\text{Property 1,}}
\end{align*}

Also, for each $i$, $  i \in \mathbb{N}_{n}$ we have:

\begin{align*}
1                                         &=\displaystyle \sum_{u\in N(y^1_i)}f(u)w(u) \\
                                          &= (1-k)f(y^2_i)+\sum_{j} \Big( f(x_{i,j})\times a_j\Big) \\
                                          &\Rightarrow \sum_{j}f(x_{i,j})=3 & {\text{Property 2,}}
\end{align*}

Now, define the  partition $A_1, \ldots , A_n$, where for each $l$, $  l \in \mathbb{N}_{ n}$, $A_l=\{a_j: f(x_{l,j})=1\}$.
By Property 1 and Property 2, $A_1, \ldots , A_n$ are disjoint and each one has three members.
Next, assume that $\{a_1, \ldots, a_{3n}\}$ can be partitioned into $n$ disjoint sets $A_1, \ldots , A_n$ such that,
for each $i$, $  i \in \mathbb{N}_{ n}$, $ \sum _{a \in A_i}a= k$. Define the function $f$ such that
for each $l$, $ l \in \mathbb{N}_{ n}$, $A_l=\{a_j: f(x_{l,j})=1\}$. By Property 1 and Property 2, $f$ is
a 1-in-Degree coloring for the graph $G$.
Therefore, $G$ has a 1-in-Degree coloring if and only if
$\{a_1, \ldots, a_{3n}\}$ can be partitioned into $n$ disjoint sets $A_1, \ldots , A_n$ such that,
for each $i$, $  i \in \mathbb{N}_{ n}$, $ \sum _{a \in A_i}a= k$. This completes the proof.

}\end{proof}

\section{The edge versions of 1-in-Degree and Not-All-Equal decompositions}\label{Section7}

An edge coloring $f:E(G)\rightarrow \{0,1\}$ for a graph $G$
is called {\it 1-in-Degree edge coloring} if and only if for every vertex $v$, $\sum_{e \ni v}f(e)=1$.
The graph $G$ has a  1-in-Degree edge coloring if and only if $G$ has a perfect matching.
Now, consider the edge version of the NAE. An edge coloring $f:E(G)\rightarrow \{red,blue\}$ of a graph $G$
is called {\it NAE edge coloring} if and only if for
 every vertex $v$, there are edges $e$ and $e'$ incident
  with $v$, such that $f(e)\neq f(e')$.

\begin{thm}\label{T6}
For a given connected graph $G$ with $\delta(G)>1$, $G$ has a NAE edge coloring if and only if $G$ is not an odd cycle.
\end{thm}

\begin{proof}{

If $G$ is an odd cycle, then the graph $G$ does not have any  NAE edge coloring. Also,
for every even cycle $\mathcal{C}$ any proper edge coloring of $\mathcal{C}$ is a  NAE edge coloring for its edges.
Let  $G$ be a connected graph with $\delta(G)\geq 2$ and  $\Delta(G)\geq 3$. Two cases can be considered:
\\
{{\bf Case 1:}}  $\Delta(G)\geq 4$. Let $v$ be a vertex of degree
more than three. The number of vertices of odd degrees is even. Let
$x$ and $y$ be two vertices  of odd degrees in $G$, put a new vertex
$t_{x,y} $ in the graph and join the vertex $t_{x,y} $ to the vertices
$x$ and $y$. Repeat this procedure until the resultant graph does not
have any vertex of odd degree. Call the resultant graph $G'$. $G'$ has
an Eulerian cycle $\mathcal{C}$. Start  from $v$ and alternatively color
the edges of $\mathcal{C}$ by $red$ and $blue$. Call this coloring $f$. Without loss of generality assume that
$\mathcal{C}=e_1e_2\ldots,e_{|E(G')|} $. By the structure of the graph $G'$ and since the degree of vertex $v$ is at least four,
for every vertex $u$, there is an index $i$, such that $e_i,e_{i+1}\in E(G)$ and the vertex $u$ is incident with
$e_i$ and $e_{i+1}$. Thus, $f(e_i) \neq f(e_{i+1})$.
Consequently, the function $f$ is a NAE edge coloring for $G$.
\\
{{\bf Case 2:}}  $\Delta(G)=3$. The proof of this case is by induction on the number of edges. There
are two vertices $v$ and $u$ such that $ d(v)=d(u)=3$ and there exists a
path $\mathcal{P}=vx_1x_2\ldots x_iu$ such that $d(x_1)=\cdots=d(x_i)=2$. Consider
the subgraph $H=G\setminus\{ x_j | j\in \mathbb{N}_i\}$ (note that $u$ and $v$ can be
adjacent, in this case we put $H=G\setminus \{e=vu\} $). $H$ contains at most two connected components.
Start  from $vx_1$ and alternatively color the edges of $\mathcal{P}$ by $red$ and $blue$. Call this partial coloring $\ell$.
First, suppose that
no connected component of $H$ is an odd cycle. By the inductive hypothesis, color each connected component of $H$ with $red$ and $blue$. The resultant coloring is an NAE edge coloring for $G$.
Next, suppose that some connected components of $H$ are odd cycles.
Note that there is an edge coloring $f:E(C_{2k+1})\rightarrow \{red,blue\}$ for the odd cycle $C_{2k+1}$
such that for every vertex $a\in V(C_{2k+1})$, except exactly one vertex $b\in V(C_{2k+1})$, there are edges $e$ and $e'$ incident with $a$, such that $f(e)\neq f(e')$. If for every edge $e$ incident with $b$, $f(e)=red$, we call this coloring, red coloring with respect to $b$. And if, for every edge $e$ incident with $b$, $f(e)=blue$, we call this coloring, blue coloring with respect to $b$.

\begin{figure}[ht]
\begin{center}
\includegraphics[scale=.8]{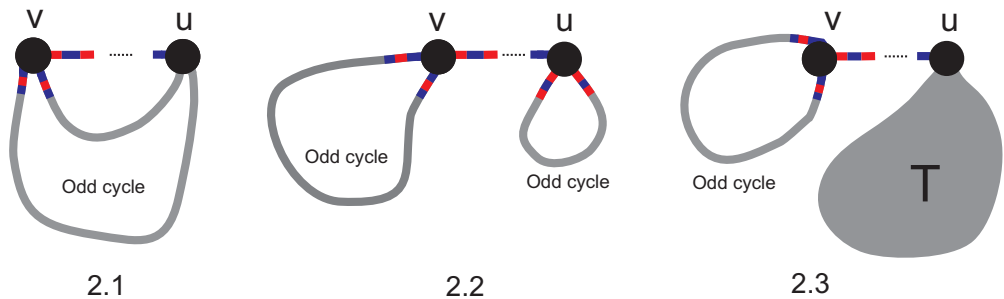}
\caption{Three subcases in the proof of Theorem \ref{T6}.}
\end{center}
\end{figure}

Now, suppose that some connected components of $H$ are odd cycles. Three subcases can be considered:
\\
{{\bf Subcase 2.1:}} If $H$ has exactly one nontrivial connected component and this connected component is an odd cycle.
Then, without loss of generality suppose that, the connected component contains $v$, consider a blue coloring with respect to $v$  for $H$. The resultant coloring is an NAE edge coloring for $G$.
\\
{{\bf Subcase 2.2:}} If $H$ has two connected components and each connected component is an odd cycle. Then, without loss of generality suppose that
$f(x_iu)=blue$. Consider a blue coloring with respect to $v$ for the connected component that contains $v$ and a red coloring with respect to $u$ for the connected component that contains $u$. The resultant coloring is an NAE edge coloring for $G$.
\\
{{\bf Subcase 2.3:}} If $H$ has two connected components and exactly
one of these  connected components is an odd cycle. Then, without loss
of generality suppose that the connected component that contains $v$ is
an odd cycle. Consider a blue coloring with respect to $v$ for the connected
 component that contains $v$ and by the inductive hypothesis, color the
 connected component of $H$ that contains $u$ with $red$ and $blue$. The resultant coloring is a NAE edge coloring for $G$.

}\end{proof}

\section{Concluding remarks}
\label{Section8}

\textbullet $ $ In this work we introduced the concept of the NAE decomposition and the concept of 1-in-Degree decomposition of graphs.
The NAE decomposition of a graph $G$ is a
 decomposition of the vertices into two sets such that each vertex in the graph $G$
has at least one
neighbor in each part. Also, the 1-in-Degree decomposition of a graph $G$ is a decomposition of the vertices of $G$
 into two sets $A$ and $B$
such that each vertex in the graph
has exactly one
neighbor in part $A$.
A summary of the results on the NAE and 1-in-Degree decompositions was shown in Table 1.
The ultimate goal of determining
the computational complexity of deciding whether a given  3-regular
bipartite graph $G$  has a NAE   decomposition, remains outstanding  for the moment.
Any hardness result can be interesting to work on.

\textbullet $ $ In the NAE and 1-in-Degree decompositions, we partition the vertices
of a graph into two parts with some properties, other types of
these problems were considered by several authors, for instance,
Gerber and Kobler introduced  the problem of deciding if a given
graph has a vertex partition into two nonempty parts such that each
vertex has at least as many neighbors in its part as in the other
part \cite{gerber2000algorithmic}. For more information about this
problem see \cite{ bazgan2006satisfactory, gerber2003algorithms, shafique2009partitioning}.

\section{Acknowledgments}
\label{}

The authors would like to thank the anonymous referees for their useful comments
 which helped to improve the presentation of this paper.

\small
\bibliographystyle{plain}
\bibliography{Dynamic}

\end{document}